\documentclass[prl,twocolumn,superscript address]{revtex4}

\usepackage{graphicx,epic,eepic,epsfig,amsmath,latexsym,amssymb,verbatim,subfigure,color}
\usepackage{verbatim}

\usepackage{theorem}

\def\squareforqed{\hbox{\rlap{$\sqcap$}$\sqcup$}}
\def\qed{\ifmmode\squareforqed\else{\unskip\nobreak\hfil
\penalty50\hskip1em\null\nobreak\hfil\squareforqed
\parfillskip=0pt\finalhyphendemerits=0\endgraf}\fi}
\def\endenv{\ifmmode\;\else{\unskip\nobreak\hfil
\penalty50\hskip1em\null\nobreak\hfil\;
\parfillskip=0pt\finalhyphendemerits=0\endgraf}\fi}

\mathchardef\ordinarycolon\mathcode`\:
\mathcode`\:=\string"8000
\def\vcentcolon{\mathrel{\mathop\ordinarycolon}}
\begingroup \catcode`\:=\active
  \lowercase{\endgroup
  \let :\vcentcolon
  }

\newcommand{\nc}{\newcommand}
\nc{\rnc}{\renewcommand}
\nc{\beq}{\begin{equation}}
\nc{\eeq}{{\end{equation}}}
\nc{\beqa}{\begin{eqnarray}}
\nc{\eeqa}{\end{eqnarray}}
\nc{\lbar}[1]{\overline{#1}}
\nc{\bra}[1]{\langle#1|}
\nc{\ket}[1]{|#1\rangle}
\nc{\ketbra}[2]{|#1\rangle\!\langle#2|}
\nc{\braket}[2]{\langle#1|#2\rangle}
\nc{\proj}[1]{| #1\rangle\!\langle #1 |}
\nc{\avg}[1]{\langle#1\rangle}
\nc{\Rank}{\operatorname{Rank}}
\nc{\smfrac}[2]{\mbox{$\frac{#1}{#2}$}}
\nc{\id}{\operatorname{id}}
\nc{\1}{\openone}
\nc{\ox}{\otimes}
\nc{\dg}{\dagger}
\nc{\dn}{\downarrow}
\nc{\cA}{{\cal A}}
\nc{\cB}{{\cal B}}
\nc{\cC}{{\cal C}}
\nc{\cD}{{\cal D}}
\nc{\cE}{{\cal E}}
\nc{\cF}{{\cal F}}
\nc{\cG}{{\cal G}}
\nc{\cH}{{\cal H}}
\nc{\cI}{{\cal I}}
\nc{\cJ}{{\cal J}}
\nc{\cK}{{\cal K}}
\nc{\cL}{{\cal L}}
\nc{\cM}{{\cal M}}
\nc{\cN}{{\cal N}}
\nc{\cO}{{\cal O}}
\nc{\cP}{{\cal P}}
\nc{\cR}{{\cal R}}
\nc{\cS}{{\cal S}}
\nc{\cT}{{\cal T}}
\nc{\cX}{{\cal X}}
\nc{\cZ}{{\cal Z}}
\nc{\supp}{{\operatorname{supp}}}
\nc{\var}{\operatorname{var}}
\nc{\rar}{\rightarrow}
\nc{\lrar}{\longrightarrow}
\nc{\polylog}{\operatorname{polylog}}

\nc{\RR}{{{\mathbb R}}}
\nc{\CC}{{{\mathbb C}}}
\nc{\FF}{{{\mathbb F}}}
\nc{\NN}{{{\mathbb N}}}
\nc{\ZZ}{{{\mathbb Z}}}
\nc{\PP}{{{\mathbb P}}}
\nc{\QQ}{{{\mathbb Q}}}
\nc{\UU}{{{\mathbb U}}}
\nc{\EE}{{{\mathbb E}}}

\nc{\be}{\begin{equation}}
\nc{\ee}{{\end{equation}}}
\nc{\bea}{\begin{eqnarray}}
\nc{\eea}{\end{eqnarray}}
\nc{\<}{\langle}
\rnc{\>}{\rangle}
\nc{\Hom}[2]{\mbox{Hom}(\CC^{#1},\CC^{#2})}
\nc{\rU}{\mbox{U}}

\begin{document}
\author{Graeme Smith}
\affiliation{Institute for Quantum Information, Caltech 107---81,
    Pasadena, CA 91125, USA}
\affiliation{IBM T.J. Watson Research Center, Yorktown Heights, NY 10598}
\author{John A. Smolin}
\affiliation{IBM T.J. Watson Research Center, Yorktown Heights, NY 10598}
\title{Degenerate Quantum Codes for Pauli Channels}

\begin{abstract}
A striking feature of quantum error correcting codes is
that they can sometimes be used to correct more errors than they can
uniquely identify.  Such {\em degenerate} codes have long been known,
but have remained poorly understood.  We provide a heuristic for
designing degenerate quantum codes for high noise rates, which is
applied to generate codes that can be used to communicate over almost
any Pauli channel at rates that are impossible for a nondegenerate
code.  The gap between nondegenerate and degenerate code performance
is quite large, in contrast to the tiny magnitude of the only previous
demonstration of this effect.  We also identify a channel for which
none of our codes outperform the best nondegenerate code and show that
it is nevertheless quite unlike any channel for which nondegenerate
codes are known to be optimal.
\end{abstract}

\date{\today} \maketitle It was Shannon \cite{Shannon48} who
discovered, by a random coding argument, the beautiful fact that the
capacity of a noisy channel $\cN$ is equal to the maximal mutual
information between an input variable, $X$, and its
image under the action of the channel:
\begin{equation}
C={\rm max}_X I(X;{\cal N}(X)) \label{Eq:ShannonCap}.
\end{equation}
It is remarkable that this maximization is over a single
input to the channel; it does not require consideration of inputs
correlated over many channel uses.  

One would hope that, as in the classical case, 
there is some measure of quantum correlations that can
be maximized over inputs to a quantum channel to give the capacity.
Unfortunately, this appears not to be the case.  The 
natural generalization of Eq.~(\ref{Eq:ShannonCap}) 
is to replace the random variable $X$ with a quantum state $\rho$
and the mutual information with the {\em coherent information} $I^{\rm c}$ 
giving
\begin{equation}\label{Eq:SingleHash}
Q_1={\rm max}_{\rho}\,I^{\rm c}({\cal N},{\rho}),
\end{equation}
where 
\begin{equation}
I^{\rm c}(\cN,\rho)=I^{\rm c}
\left(I{\otimes}{\cal N}(\proj{\phi^{AB}})\right)\ .
\end{equation}
Here $\ket{\phi_{AB}}$ is a purification of $\rho$.  Its use reflects
the fact that unlike in the classical case, there can be no remaining
copy of the channel input with which to compare correlations--instead
we must consider the quantum state as a whole.  The coherent information
is defined by $I^{\rm c}(\rho_{AB})=S(\rho_B){-}S(\rho_{AB})$ with 
$S(\rho){=}-{\rm Tr}(\rho\log \rho)$.

While we can achieve $Q_1$ using a random code on the typical subspace
of the maximizing $\rho$, it has long been known that this rate is
not always optimal \cite{SS96,DSS98}.  They exhibit codes with rates
larger than $Q_1$ for very noisy depolarizing channels which have
$Q_1$ small or even zero.

The correct quantum capacity formula is not $Q_1$, but instead is given by
\cite{D03,Shor02,Lloyd97}
\begin{equation}\label{Eq:QuantumCapacity}
Q=\lim_{n\rightarrow\infty} \frac{1}{n} {\rm max}_{\rho_n} 
I^{\rm c}\left({\cal N}^{\ox n}, {\rho_n}\right), 
\end{equation}
where taking the limit $n\rightarrow\infty$ means that we must
consider the behavior of the channel on inputs entangled across many
uses.  This {\em multi-letter} formula is an expression of our
ignorance about the structure of capacity achieving codes for a
quantum channel.

The difference between these single- and multi-letter formulas is
intimately related to the existence of degenerate quantum codes.
Strictly speaking, degeneracy is not a property of a quantum code
alone, but a property of a code together with a family of errors it is
designed to correct.  More formally, one usually says that a code
$\cC$ degenerately corrects a set of errors $\cE$ if in addition to
correcting $\cE$, there are multiple errors in $\cE$ that are mapped
to the same error syndrome.  In the context of probabilistic noise,
which we will be concerned with exclusively, we say that a code $\cC$
degenerately corrects the noise due to a channel $\cN$ if it can be
decoded with a high fidelity and furthermore multiple errors in the
set of typical errors of $\cN$ are mapped to the same error syndrome.
For the most part, we will be concerned with {\em grossly
degenerate} codes, which have the further property that the number of
typical errors mapped to each syndrome is exponential in the code's
block-length.

For the depolarizing channel considered in \cite{SS96, DSS98}, as well
as the Pauli channels considered below, $Q_1$ is exactly the maximum
rate achievable with a nondegenerate code.  That $Q>Q_1$ is then
established by the construction of a massively degenerate code.  While
this was accomplished in the work of \cite{SS96,DSS98}, the difference
found was over a minuscule range of noise parameters and extremely
small in magnitude. As a result, one may have thought that
Eq.~(\ref{Eq:SingleHash}) is ``essentially correct'', with some minor
modifications in the very noisy regime.  In the decade since the
appearance of these two works, there has been almost no progress on
understanding the difference between the single- and multi-letter
expressions above, a failure which has to some extent been tempered by
the hope that the smallness of the effect would make it amenable to a
perturbative analysis.  We will show that this cannot be the case
and in fact that the smallness of the effect found in \cite{SS96,DSS98} is 
more accidental than fundamental.

Until now, very little has been understood about why the degenerate
codes of \cite{SS96,DSS98} work, besides that they seem to be highly
degenerate.  The main contribution of this paper is to provide a
conceptual explanation of why degenerate codes of this type work,
along with a related heuristic for designing codes for more general
channels.  Using this heuristic, we find better codes for almost all
Pauli channels, and exhibit cases where the effect of degeneracy can be
quite large.  This large gap between the performance of nondegenerate
and degenerate codes implies that a perturbative approach to is
unlikely to be useful.

A secondary contribution we believe to be no less important, but
which lies on the periphery of the current work, is the identification
of the two Pauli channel as an important piece of the degenerate
coding puzzle.  This channel derives no benefit from the degenerate
codes we study, but is also quite different from any of the degradable
channels, a set of channels including the dephasing and erasure
channels, and comprising the only channels for which which
nondegenerate codes are known to be optimal \cite{DS03}.  Therefore
either there is some other sort of degenerate code that {\em will}
beat $Q_1$ or $Q_1$ can be optimal for nondegradable channels.
Either outcome seems plausible, and progress on resolving this
dichotomy would necessarily deepen our understanding of the quantum
coding problem in general.  Along a similar line, we have introduced a
general method for showing that a channel is not degradable, taking
one of the first steps in the program to classify all degradable
channels.

\noindent {\it Cat Codes for Pauli Channels.}---The codes we will
consider are $m$-qubit repetition code, sometimes called a ``cat
codes'' because the code space is spanned by $\ket{0}^{\ox m}$ and
$\ket{1}^{\ox m}$.  These have stabilizers $Z_1Z_2,Z_1Z_3,\dots
Z_1Z_m$ and logical operators $\bar{X} = X^{\ox m}$ and $\bar{Z} =
Z_1$, so that an error of the form $X^{\mathbf{u}}Z^{\mathbf{v}}$
leads to syndrome $\{u_1\oplus u_2,\dots,u_1\oplus u_m\}$ and in the
absence of a recovery operation gives a logical error of
$\bar{X}^{u_1}\bar{Z}^{\oplus_l v_l}$.  By encoding half of
$\ket{\phi^{+}}_{AB} = (\ket{00}+\ket{11})/\sqrt{2}$ in our repetition
code, we get the state for which the coherent information in
Eq.~(\ref{Eq:QuantumCapacity}) will be more than $m$ times $Q_1$.
Sending the $B$ system of the resulting $\ket{\phi^{+}_m}_{AB}$
through ${\cN_{\mathbf p}}^{\!\ox m}$ and subsequently measuring the
stabilizers $\{Z_1Z_{l}\}_{l=2}^m$ leads to the state $\rho_{A\!B_m} =
\sum_{{\mathbf r} \in \{0,1\}^{m-1}} \Pr({\mathbf r}) I\ox
\cN^{{\mathbf r}}(\proj{\phi^+})\ox \proj{\mathbf r}$, where ${\mathbf
r}$ is the syndrome measured, $\cN^{\mathbf r}$ is the induced channel
given ${\mathbf r}$ (which is also a Pauli channel), and $\Pr({\mathbf
r})$ is the probability of measuring ${\mathbf r}$.  Concatenating our
repetition code with a random stabilizer code allows communication
with high fidelity at a rate of
\begin{equation}\label{Eq:CatCodeRate}
\frac{1}{m}I^{\rm c}(\rho_{A\!B_m}) = \frac{1}{m}\sum_{\mathbf r}
\Pr({\mathbf r})I^{\rm c}(I\ox \cN^{\mathbf r}(\proj{\phi^+})).
\end{equation}

Because the repetition code is highly symmetric we can find
explicit formulas for both $\Pr({\mathbf r})$ and $\cN^{\mathbf r}$,
and thus a fairly compact expression for $I^{\rm c}(\rho_{A\!B_m})$.
The joint probabilities of logical errors
and syndrome outcomes are%, as proved in the Appendix, given by 
%the following equation, which we prove in an Appendix.
\begin{widetext}
\begin{equation}\label{Eq:CatCodeProbs}
\Pr(\bar{X}^u\bar{Z}^v\!,{\mathbf r}) = \frac{1}{2}
\!\left((p_x\!{+}p_y)^{u(m{-}2r){+}r}(1{-}p_x\!{-}p_y)^{(1{-}u)(m{-}2r){+}r}\!
\!+\!(-1)^v\!(p_x\!{-}p_y)^{u(m{-}2r){+}r}(1{-}p_x
\!{-}p_y\!{-}2p_z)^{(1{-}u)(m{-}2r){+}r}\right),
\end{equation}
\end{widetext}
where $r = |{\mathbf r}|$, the Hamming weight of ${\mathbf r}$. 
Eq. (\ref{Eq:CatCodeProbs}) allows us to find both $\Pr({\mathbf r})$ and  
%$ = \binom{m-1}{r}\left(\left(p_x{+}p_y\right)^r
%\left(1{-}p_x{-}p_y\right)^{m{-}r} + 
%\left(p_x{+}p_y\right)^{m{-}r}\left(1{-}p_x{-}p_y\right)^{r}\right)$ and
the error probabilities of the induced channels $\cN^{\mathbf r}$.
%A salient feature of 
This formula depends on $r$ but has no other dependence on $\mathbf
r$, which dramatically reduces the number of induced channels that
need to be considered.

\begin{figure}[htbf]
\centering
\includegraphics[width=8cm]{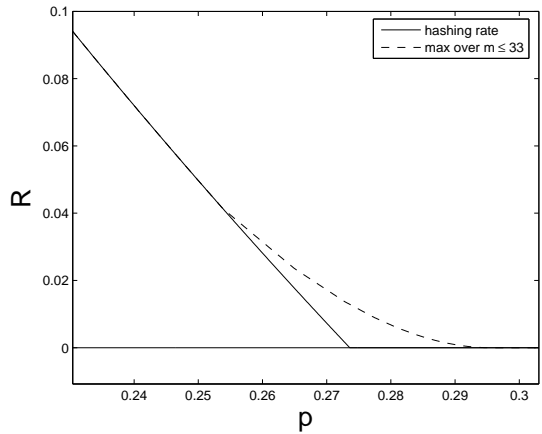}
\caption{Best $Z$-cat code rates for independent phase and amplitude
errors with $q_z/q_x=(p_z+p_y)/(p_x+p_y)=9 $ (and where $p=p_x{+}p_y{+}p_z$).
The optimal $m$ increases with $p$. $m=33$ gives the best threshold of
$\approx .295$, compared to a hashing threshold less than $.274$. The
rule of thumb $m\approx 1/p_z$ gives an estimated $m=36$, not far from
the optimum.}
\label{figure:bestRates}
\end{figure}

By evaluating (\ref{Eq:CatCodeRate}) on the probabilities of
(\ref{Eq:CatCodeProbs}), we find that for almost all Pauli channels
there is a repetition code with nonzero rate at the hashing
point.  When $p_x\gg p_z$ the best code is in the $Z$
basis with length scaling like ${1}/{p_z}$, which we'll study in
detail in the next section.  For $p_x\geq p_z\geq p_y$ it
is a good rule of thumb to use a $Z$ repetition code of length
$m\approx {1}/{p_z}$, with the largest increase in rate for fairly
asymmetrical channels (Fig. \ref{figure:bestRates}).

\noindent {\it Repetition Lengths for Almost Bitflip
Channels.}---\label{Sec:BitFlip}To illustrate the tradeoff determining
the best repetition code length, we will study their performance for
channels with independent phase and amplitude error probabilities.  An
error $X^uZ^v$ is said to be a phase error if $v=1$ and an amplitude
error if $u=1$ (note that when $u=1$ and $v=1$ it is both).
Throughout, we define $q_x{=}p_x{+}p_y$ and $q_z{=}p_z{+}p_y$ to be
the amplitude and phase error probabilities, respectively, and in a
slight abuse of terminology refer to amplitude and phase errors as $X$
and $Z$ errors, with a $Y$ error being ``both $X$ and $Z$.''
Independence of phase and amplitude errors requires $p_x =
q_x(1{-}q_z)$, $p_y=q_xq_z$, and $p_z = q_z(1{-}q_x)$.  When $q_x \gg
q_z$ we find that the repetition code with the best zero-rate noise
threshold has $m\approx 1/q_z$, which can be understood by considering
the effective channels induced by the code.

The independence of phase and amplitude, together with our
generators involving only $Z$'s tells us that the probability 
of a logical phase error is independent of $\mathbf r$, and given by
$q_{\bar{Z}}{=}\Pr\left(\oplus_{l=1}^m v_l{=}1\right)=[1{-}(1{-}2q_z)^m]/2$,
which also follows from Eq. (\ref{Eq:CatCodeProbs}).

As we have already seen, the probability of a logical amplitude error
depends only on $r=|{\mathbf r}|$, not on $\mathbf r$ itself.  If $m$
is large, the probability distribution of $r$ becomes concentrated
near $r_o\equiv (m{-}1)q_x$ and $r_1\equiv (m{-}1)(1{-}q_x)$.  This is
because there are typically $(m{-}1)q_x$ $X$ errors on qubits $2$
through $m$ and these qubits all get flipped if qubit $1$ has an $X$
error.  So, the measured value of $r$ tells us whether or not a
logical $X$ error has occurred, at least with high probability.  One
can see from this, together with the $q_{\bar{Z}}$ above, that as $m$
increases we learn more about the logical $X$ error at the expense of
knowing less about the logical $Z$.

Indeed, the optimal repetition length will minimize the entropy in the
logical qubits conditioned on $r$, which near the hashing point occurs
when the repetition length is around $1/q_z$, at which point almost
all of the $X$ entropy has been removed.  If we increase $m$ beyond
this the gain in information about the logical $X$ is less than the
resulting decrease in our knowledge of the logical $Z$'s.  The overall
rate thus achieved at the hashing point is ${2q_z\ln(1/q_z)}/{\ln \ln
(1/q_z)}$.

Note that essentially all of the entropy in the $X$ errors is removed
by the best code, with the optimal length determined by a tradeoff
between the reduction of entropy in the $X$ errors and the increase of
entropy in the $Z$ errors. This sort of tradeoff also determines the
optimal repetition code length for a general Pauli channel.

\begin{figure}[htbf]
\centering
\includegraphics[width=8cm]{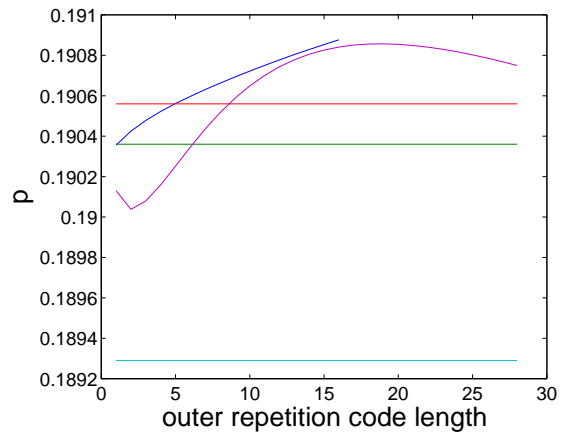}
\caption{Error probability where the rate goes to zero, as a function
of length of second level cat code.  Here the horizontal axis is $m$,
the length of the second level cat code.  The bottom line is hashing,
the middle line is a 5 qubit repetition code, the upper line is a
concatenated 5 in 5 repetition code. The lower curve is a 3 in $m$
repetition code; the upper is 5 in $m$. }
\label{Fig:Depolar}
\end{figure}

\noindent{\it Concatenated Repetition Codes.}---\label{Sec:ConCat} We can
immediately apply this analysis to design even better codes by using
concatenation.  By adapting a second level of repetition code to the
error probabilities of the channels induced by the first level we can
exceed the performance of any single level cat code.  We have used
this approach for the depolarizing channel with the results shown in
Fig.~\ref{Fig:Depolar}, where we plot the probabilities at which the
rate of a concatenated $3$ in $m$ and $5$ in $m$ code goes to zero as
a function of $m$, the size of the outer cat code. If we first use a
3-cat code in the $Z$ basis, followed by an $m$-cat code in the $X$
basis, we find the highest threshold for a $3$ in $19$ code, with a
nonzero rate up to $p \approx 0.19086$, surpassing the codes of
\cite{DSS98}.  Starting with a 5-cat code the threshold increases up
to $p\approx 0.19088$ for $m=16$, the best known code for this
channel, but for higher values of $m$ the computation of this
probability is quite slow. Based on the character of the channels
induced by the inner repetition code, together with the behavior for
$m \leq 16$ we expect that the threshold increases until something
like $5$ in $25$, at which point a larger $m$ begins to reduce the
effectiveness of the code.

\noindent{\it Two-Pauli Channels are Special.}---Besides the one-Pauli
channels, the only channels for which we can find no code offering an
advantage near the hashing point are tightly concentrated near
$\cN^{\rm tp}_{p}(\rho)\equiv (1-p)\rho{+}\frac{p}{2}X\rho
X{+}\frac{p}{2}Z\rho Z$.  While hashing is optimal for one-Pauli 
channels \cite{Rains99}, $\cN^{\rm tp}_{p}$ is not
known to have additive coherent information, which is equivalent to
the optimality of hashing.  Furthermore, we will show that unlike all
channels known to be additive this channel is not degradable
\cite{DS03}.

Every channel $\cN$ can be expressed as an isometry followed by a
partial trace, which is to say there is an isometry
$U_{\cN}:{A}\rightarrow{BE}$ such that $\cN(\rho){=}{\rm
Tr}_{E}U_{\cN}\rho U_{\cN}^\dg$.  The complementary channel of $\cN$,
called $\cN^C$, results by tracing out system $B$ rather than $E$:
$\cN^C(\rho){=}{\rm Tr}_{B}U_{\cN}\rho U_{\cN}^\dg$.  A channel is
called degradable if there is a completely positive map,
$\cD:B\rightarrow E$, which ``degrades'' $\cN$ to $\cN^C$, so that
$\cD\circ \cN{=}\cN^C$.  The existence of such a map immediately
implies the additivity of $I^{\rm c}$\cite{DS03}, which can be seen by
noting that $I^{\rm c}(\cN^{\otimes(n_1{+}n_2)},
\rho_{n_1n_2}){\leq}I^{\rm
c}(\cN^{\otimes n_1},\rho_{n_1}){+}I^{\rm c}(\cN^{\otimes n_2},
\rho_{n_2})$ exactly when $I(E_{n_1};E_{n_2}) \leq
I(B_{n_1};B_{n_2})$ and recalling that $I(B_{n_1};B_{n_2})$ cannot
increase under local operations.  We now show there is no such $\cD$
for $\cN^{\rm tp}_{p}$ when $0<p<1$.

Letting $\cN^{\rm tp}_{p}(\ketbra{i}{j}) = 
\sum_{kl}N_{ij;kl}\ketbra{k}{l}$ define $N$ and 
${{\cN^{\rm tp}_{p}}^C(\ketbra{i}{j})} = 
\sum_{kl}N_{ij;kl}^C\ketbra{k}{l}$ define  $N^C$,  
we find
\begin{equation}\nonumber
N = \left(\begin{matrix}
1{-}{p}/{2}&0&0&{p}/{2}\\
0&1{-}{3p}/{2}&{p}/{2}&0\\
0&{p}/{2}&1{-}{3p}/{2}&0\\
{p}/{2}&0&0&1{-}{p}/{2}
\end{matrix}\right) {\rm and}
\end{equation}
\begin{equation}\nonumber
N^C =\left(\begin{matrix}
{p}/{2} & 0 & 0 & 0 & {p}/{2} & \alpha & 0 & \alpha& 1{-}p \\
0 & {-p}/2 & \alpha  & {p}/2 & 0 & 0 & \alpha & 0 & 0 \\ 
0 & {p}/{2} & \alpha & {-p}/{2} & 0 & 0 & \alpha & 0 & 0 \\  
{p}/2 & 0 & 0 & 0 & {p}/{2} & -\alpha & 0 & -\alpha & 1{-}p
\end{matrix}\right),
\end{equation}
where $\alpha=\sqrt{p(1-p)/{2}}$.
If $\cN^{\rm tp}_{p}$ is degradable, there must be a CPTP map $\cD$ such 
that $\cD\circ\cN = \cN^C$, which is equivalent
to $ND = N^{C}$, with $D$ defined by $\cD(\ketbra{s}{t}) = 
\sum_{uv}D_{st;uv}\ketbra{u}{v}$.  For $N$ and $N^C$ as above, this gives
\begin{equation}\nonumber
D = \left(\begin{matrix}
{p}/{2} & 0 & 0 & 0 & {p}/{2} & \beta & 0 & \beta & 1{-}p\\
0 & -\gamma & \beta & \gamma & 0 & 0 & \beta & 0 & 0\\
0 & \gamma & -\beta & \gamma & 0 & 0 & \beta & 0 & 0\\
{p}/{2} & 0 & 0 & 0 & {p}/{2} & -\beta & 0 & -\beta & 1{-}p
\end{matrix}\right)
\end{equation}
with $\beta=\sqrt{p/(2-2p)}$ and $\gamma={p}/(2-4p)$.
The Choi matrix \cite{Choi75} of $\cD$, $C^{\cD}_{ik;jl} = D_{ij;kl}$, is thus
\begin{equation}\nonumber
C^{\cD} = \left(\begin{matrix}
{p}/{2} & 0 & 0 & 0 & -\gamma & \beta\\
0 & {p}/{2} & \beta & \gamma & 0 & 0 \\
0 & \beta & 1{-}p &  \beta & 0 & 0 \\
0 & \gamma & \beta & {p}/{2} & 0 & 0 \\
-\gamma & 0 & 0 & 0 & {p}/{2} & -\beta\\
\beta & 0 & 0 & 0 & -\beta & 1{-}p
\end{matrix}\right)
\end{equation}
which contains the subblock 
$\binom{p/2\ -\gamma}{-\gamma\ p/2}$.  This has a negative eigenvalue
for all $0<p<1$, so that $C^{\cD}$ cannot be nonnegative and thus
$\cD$ is not CP.

Besides repetition codes, we have explored concatenated repetition
codes for $\cN^{\rm tp}_{p}$, all of which performed worse than the
hashing rate of $1{-}H(p){-}p$.  This suggests the capacity of
$\cN^{\rm tp}_{p}$ is exactly $1{-}H(p){-}p$, and in light of its
nondegradability we hope a proof of this conjecture will point towards
a new sufficient criterion for the additivity of coherent information.

\noindent{\it Discussion.}---It is tempting to ask if there is a simpler
characterization of the quantum channel capacity than is provided
by Eq.~(\ref{Eq:QuantumCapacity}).  In particular, contrary to what is
sometimes claimed, the results of \cite{DSS98,SS96} and this work {\em
do not rule out a single letter formula for the capacity}---what is
ruled out is the possibility that the single letter optimized coherent
information is the correct formula.  It could be that there {\em is} a
single letter formula for the capacity, or less ambitiously simply an
efficiently calculable expression, which takes degeneracy into
account.  The characterization of capacity in terms of coherent
information is fundamentally nondegenerate, and it may be this which
leads to the necessity of regularization, rather than an inherent
superadditivity of quantum information.

%On a similar note, it would be nice to find families
%of quantum codes that are capacity approaching with high probability.
%This is not the case for random
%stabilizer codes, nor for random codes on the typical subspace of a
%state maximizing the single letter coherent information, but perhaps
%by explicitly considering the codes' degeneracy
%progress could be made.  

More concretely, the two-Pauli channel with equal probabilities 
seems to be somehow different from other Pauli
channels.  Given their success with almost all other Pauli channels,
the failure of cat codes to beat $Q_1$ in
this case suggests that hashing is optimal.
Resolving this conjecture seems to be a manageable problem
whose solution may lead to a better understanding of
additivity questions for quantum channels in general.

The ideas explored here are also useful
for quantum key distribution.  In particular, using highly structured codes for
information reconciliation improves the noise threshold
of BB84 with one-way classical post-processing from $12.4\%$ to
$12.9\%$ \cite{SRS06}.
    
Finally, we hope the coding approach suggested by the almost bitflip
channel
%---managing the tradeoff between the reduction of entropy in logical 
%amplitude errors and the increase of entropy in logical phase errors---
will 
lead to codes with rates beyond what we have presented here.
Focusing on reducing the amplitude error rate with an inner code while
trying to avoid scrambling the phase errors more than necessary and
following this up with a random stabilizer code (or perhaps a second
similarly chosen code reversing the roles of amplitude and phase)
offers an appealing heuristic for code design.  Viewed in this way,
the inner codes we have considered are quite primitive---a repetition
code is the simplest code there is---and it seems likely more
sophisticated codes will perform better.

In summary, we have provided a toolset for studying degenerate
codes on Pauli channels.  We have demonstrated channels and codes for 
which the gap between the degenerate and nondegenerate performance is quite 
large compared to previous results, and improved the threshold for the
more generally applicable depolarizing channel.  Whether the capacity of 
the two-Pauli channel can be improved by degenerate codes remains an
open question, the solution to which will likely prove illuminating.
 
We are grateful to DP DiVincenzo, D
Leung, and MB Ruskai for valuable discussions.  GS thanks NSF
PHY-0456720 and NSERC of Canada. JAS thanks ARO contract
DAAD19-01-C-0056.

%\bibliographystyle{apsrev}
%\bibliography{Sausage}
\end{document}